\documentclass[prb,twocolumn,showpacs]{revtex4}
\usepackage{color,soul}
\usepackage{amsmath}
\usepackage{dcolumn}
\usepackage{graphicx}

\begin{document}

\title[Short title for running header]{Detecting the orbital character of the spin fluctuation in the Iron-based superconductors with the resonant inelastic X-ray scattering spectroscopy}
\author{Da-Wei Yao and Tao Li}
\affiliation{Department of Physics, Renmin University of China, Beijing 100872, P.R.China}
\date{\today}

\begin{abstract}
The orbital distribution of the spin fluctuation in the iron-based superconductors(IBSs) is the key information needed to understand the magnetism, superconductivity and electronic nematicity in these multi-orbital systems. In this work, we propose that the resonant inelastic X-ray scattering(RIXS) technique can be used to probe selectively the spin fluctuation on different Fe $3d$ orbitals. In particular, the spin fluctuation on the three $t_{2g}$ orbitals, namely, the $3d_{xz}$, $3d_{yz}$ and the $3d_{xy}$ orbital, can be selectively probed in the $\sigma\rightarrow\pi'$ scattering geometry by aligning the direction of the outgoing photon in the $y$, $x$ and $z$ direction. Such orbital-resolved information on the spin fluctuation is invaluable for the study of the orbital-selective physics in the IBSs and can greatly advance our understanding on the relation between orbital ordering and spin nematicity in the IBSs and the orbital-selective pairing mechanism in these multi-orbital systems.
\end{abstract}


\maketitle
The multi-orbital nature of the iron-based superconductors(IBSs) is the most important origin for the novel properties of this new family of unconventional superconductors. Mounting evidences have been accumulated through the years for the importance of the orbital degree of freedom in the magnetism, superconductivity and the electronic nematicity of the system. For example, electron in different Fe $3d$ orbitals may experience different strengths of electron correlation and a two component picture with both itinerant electron and local moment may be necessary to understand the magnetism of the IBSs\cite{Zhang,Kou,Shen,Si,Capone}. Indeed, recent ARPES measurements have found large difference in the mass renormalization factor for electron in the five Fe $3d$ orbitals. Such a phenomena is generally termed as orbital-selective Mottness(OSMT). At the same time, electron in different Fe $3d$ orbitals may prefer different magnetic correlation patterns\cite{Lee,Dagotto,Nevid,Su1,Su2}. More specifically, electron in the $3d_{yz}$ and $3d_{xz}$ orbital prefer magnetic order with a wave vector of $\mathrm{Q}=(\pi,0)$ and $\mathrm{Q'}=(0,\pi)$ respectively as a result of their different nesting condition on the Fermi surface. Such a  phenomena is dubbed orbital-selective spin fluctuation(OSSF) by Fanfarillo \textit{et al} \cite{Fanfarillo1,Fanfarillo2,Fanfarillo3,Benfartto}. The study of the origin and nature of the OSMT and OSSF phenomena is the key to understand the relation between orbital ordering and spin nematicity in the IBSs and the orbital selective pairing mechanism in these multi-orbital systems\cite{Luo,Beak,Kuroki,LeeDH,Kotliar,Kreisel,Sprau}. 

With such a rich expectations on the orbital-selective physics in the magnetism of the IBSs, it is quite embarrassing to find that there is almost no efficient way to probe the orbital character of the spin fluctuation in the IBSs. In principle, the orbital character of the magnetic fluctuation can be inferred from the atomic form factor in the inelastic neutron scattering measurement. However, the difference in the atomic form factor of the Fe $3d$ orbitals is rather small, since the atomic form factor is determined solely by the electron density distribution, rather than the wave function of the electron orbital. 

\begin{figure}
\includegraphics[width=8cm]{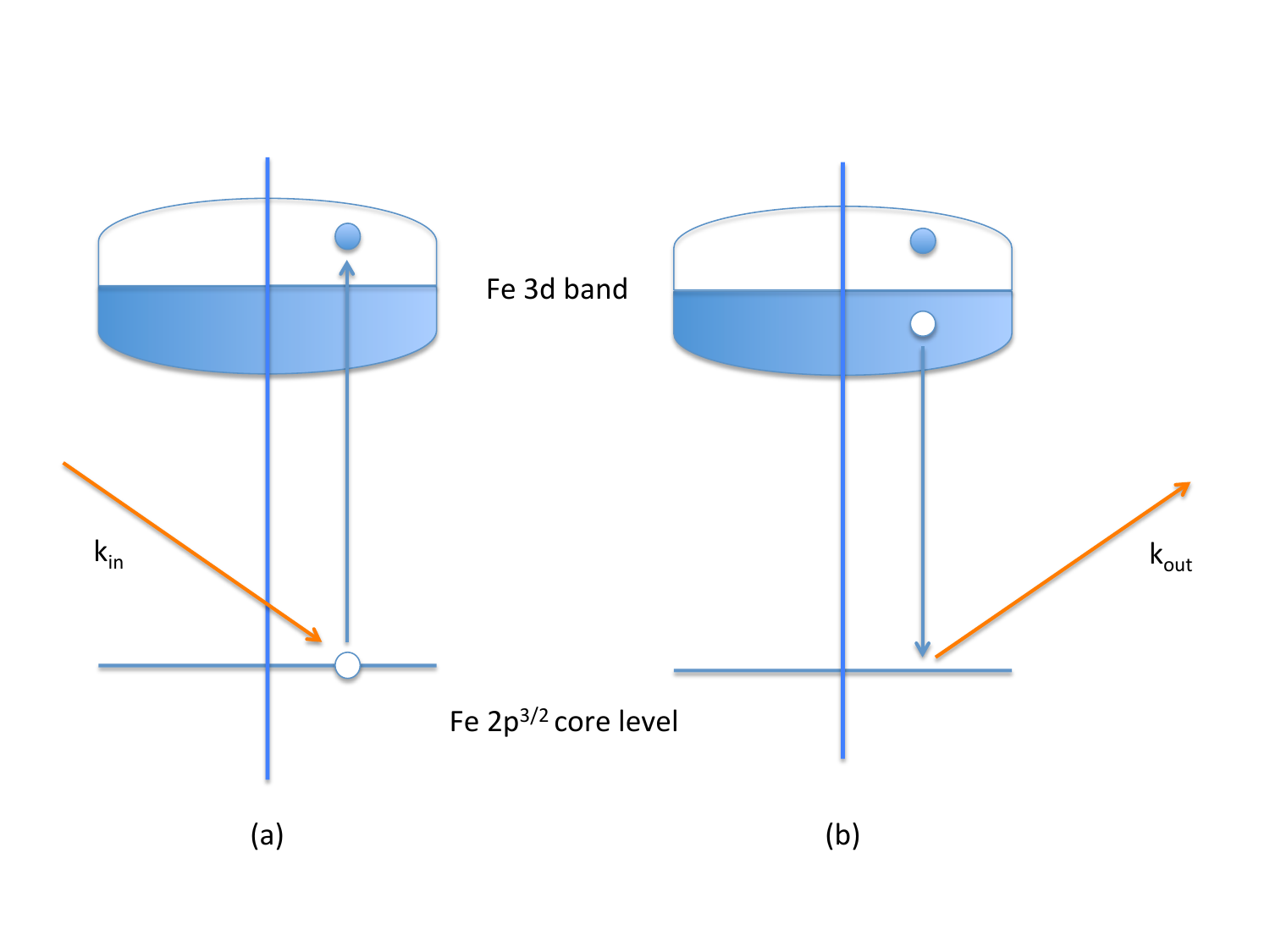}
\caption{\label{fig1}
(Color on-line) Illustration of the Fe $L_{3}$-edge RIXS process studied in this work. The Fe $L_{3}$-edge RIXS process is a direct RIXS process and is composed of the following two steps: (a)the incident photon with momentum $\mathrm{k}_{in}$ excites a Fe $2p^{\frac{3}{2}}$ core electron into the Fe $3d$ shell, (b)the Fe $2p^{\frac{3}{2}}$ core hole left behind by the photon excitation is annihilated by another electron in the Fe $3d$ shell, accompanied by the emission of a scattered photon with momentum  $\mathrm{k}_{out}$. The RIXS process effectively creates a particle-hole pair in the Fe $3d$ shell.}
\end{figure}

In this paper, we propose that the resonant inelastic X-ray scattering(RIXS) technique can be used to probe the orbital character of the spin fluctuation in the IBSs. We find that one can couple selectively to the spin fluctuation on a given Fe $3d$ orbital by choosing properly the polarization of the incident and outgoing photon. We have derived the explicit form for the RIXS selection rule and suggested the relevant scattering geometry for studying the OSMT and OSSF physics in the IBSs. The information provided by such orbital-selective measurement will greatly advance our understanding on the relation between orbital ordering and spin nematicity in the IBSs and the orbital-selective pairing mechanism in these multi-orbital systems.

In this study, we will focus on direct RIXS process, which is a two-step process to probe the electronic excitation in the valence shell using an atomic resonance between a core level and the valence shell as an intermediate step\cite{RIXS}. The Fe $L_{3}$-edge resonance between the Fe $2p^{\frac{3}{2}}$ core level and the Fe $3d$ valence shell is just such a resonance and has already been used to probe the magnetic fluctuation in the heavily doped IBSs\cite{Zhou}. The RIXS process at the Fe $L_{3}$-edge is illustrated in Fig.1. The incident photon first excite a core electron in the Fe $2p^{\frac{3}{2}}$ state into the Fe $3d$ shell(Fig.1a). The hole left behind by photon excitation is then annihilated by another Fe $3d$ valence electron(Fig.1b), accompanied by the emission of a scattered photon. The RIXS process thus effectively creates a particle-hole pair in the Fe $3d$ shell. As the resonance energy can be quite high, the momentum transfer by the scattered photon can be quite large. RIXS can thus be used to probe excitation in a large portion of the Brillouin zone. At the same time, as a result of the strong spin-orbital coupling in the Fe $2p^{\frac{3}{2}}$ core level, RIXS can also be used to probe the spin dynamics of the valence electron. We note that the Fe $L_{3}$-edge RIXS spectrum of the IBSs has already been computed in the random phase approximation scheme in a previous study\cite{Tohyama}. However, the power of polarization resolution has not been fully explored in this early study, in which only the polarization of the incident photon is specified. This greatly limit the possibility of resolving the orbital character of the spin fluctuation in the IBSs with the RIXS technique.

The general theory of direct RIXS process is now well established\cite{RIXS0} and has been summarized in the excellent review article\cite{RIXS} by Ament \textit{et al}. Here we will adopt the notations of this review. Using the second order perturbation theory, the transition amplitude induced by the photon scattering can be expressed as\cite{RIXS}
\begin{equation}
F_{f,g}=\langle f|\mathcal{D}_{out}^{\dagger}G(z)\mathcal{D}_{in}|g\rangle.
\end{equation}
Here $|g \rangle$ and $|f\rangle$ are the initial and final state of the system. $G(z)=\sum_{n}\frac{|n\rangle\langle n |}{z-E_{n}}$ is the propagator for the intermediated state after the first photon excitation. $|n\rangle$ is the intermediate state with a core hole, $z=E_{g}+\hbar\omega+i\Gamma_{n}$. $E_{g}$ and $E_{n}$ are the energies of the ground state and the intermediate state, $\Gamma_{n}$ is the inverse life time of the intermediate state caused by other non-radiative atomic processes. $\hbar\omega$ is the energy of the incident photon. $\mathcal{D}=\frac{1}{im\omega}\sum_{i}e^{i\mathrm{k}\cdot\mathrm{r}_{i}}\epsilon\cdot \mathrm{p}_{i}$ is the transition operator corresponding to the nonmagnetic coupling between the photon and the electron(this is just the usual $\mathrm{A}\cdot\mathrm{p}$ term). Here $\epsilon$ is the polarization vector of the incident or outgoing photon. $\mathrm{r}_{i}$ and $\mathrm{p}_{i}$ are the position and momentum operator of the $i$-th electron.

The expression of the transition amplitude $F_{f,g}$ can be simplified by the following two approximations. Firstly, since the radius of the Fe $3d$ orbital is much smaller than the distance between neighboring Fe ions in the IBSs, the dipole approximation is applicable\cite{RIXS} and we can approximate the plane wave factor $e^{i\mathrm{k}\cdot\mathrm{r}_{i}}$ in the transition operator $\mathcal{D}$ by a constant $e^{i\mathrm{k}\cdot\mathrm{R}_{i}}$, in which $\mathrm{R}_{i}$ is the position of the atomic site that the $i$-th electron resides. Under the dipole approximation, the RIXS transition amplitude becomes
\begin{equation}
F_{f,g}=\langle f|(\epsilon_{out}\cdot \mathrm{D}^{\dagger}_{\mathrm{k}_{out}})G(z)(\epsilon_{in}\cdot\mathrm{D}_{\mathrm{k}_{in}})|g\rangle,
\end{equation}
in which $\mathrm{D}_{\mathrm{k}}=\sum_{i}e^{i\mathrm{k}\cdot\mathrm{R}_{i}} \mathrm{r}_{i}$ is the dipole operator at the photon wave vector $\mathrm{k}$. Here we have used the operator identity $\mathrm{p}_{i}=\frac{m}{-i\hbar}[\frac{\mathrm{p}_{i}^{2}}{2m}+V(\mathrm{r}_{i}),\mathrm{r}_{i}]=\frac{m}{-i\hbar}[H_{i},\mathrm{r}_{i}]$ to rewrite the matrix element of $\mathcal{D}$ as 
\begin{equation}
\langle n|\mathcal{D}|g\rangle=\frac{E_{n}-E_{g}}{\hbar\omega}\langle n|\sum_{i}e^{i\mathrm{k}\cdot\mathrm{R}_{i}}\epsilon\cdot \mathrm{r}_{i}|g\rangle,\nonumber
\end{equation}
 and then used the resonance condition $\hbar\omega=E_{n}-E_{g}$. As a second approximation, we neglect the interaction effect in the intermediate state and treat the denominator of $G(z)$, namely, $z-E_{n}$, as a constant. This is the so called fast-collision approximation\cite{LuoJ,Veenendaal}, which is valid for direct RIXS process studied in this paper\cite{indirect}. Under the fast-collision approximation, $G(z)$ of the Fe $L_{3}$-edge RIXS becomes
\begin{equation}
G(z)\approx\frac{1}{i\Gamma_{2p^{\frac{3}{2}}}}\sum_{i,m_{j}}|i,\frac{3}{2},m_{j} \rangle \langle i, \frac{3}{2},m_{j} |.\nonumber
\end{equation}
Here $|i,\frac{3}{2},m_{j}\rangle$ denotes the state in which a core hole is created by the photon excitation at site $i$ in the $m_{j}$-th component of the Fe $2p^{\frac{3}{2}}$ multiplet.

Under these two approximations, one find that
\begin{equation}
F_{f,g}\propto\langle f|\ \sum_{i}e^{i\mathrm{q}\cdot\mathrm{R}_{i}}T_{i}\ |g\rangle,
\end{equation}
in which $\mathrm{q}=\mathrm{k}_{in}-\mathrm{k}_{out}$ is the momentum transfer from the scattered photon.  $T_{i}$ is a transition operator acting on the Fe $3d$ electrons on site $i$ and is given by
\begin{equation}
T_{i}=\sum_{m_{j}}r_{out}|i,\frac{3}{2},m_{j} \rangle\langle i,\frac{3}{2},m_{j} |r_{in},
\end{equation}
in which $r_{in}=\epsilon_{in}\cdot\mathrm{r}$ and $r_{out}=\epsilon_{out}\cdot\mathrm{r}$. As a result of the spin-orbital coupling in the $2p$ core level, $T_{i}$ can induce both spin conserving and spin flip transitions in the Fe $3d$ shell. Completing the summation over $m_{j}$, we find that $T_{i}$ can be written as the following second quantized form
\begin{eqnarray}
T_{i}&=&\sum_{\mu,\nu,\sigma}n_{\mu,\nu}c^{\dagger}_{i,\mu,\sigma}c_{i,\nu,\sigma}\nonumber\\
     &+&\sum_{\mu,\nu,\sigma}s^{z}_{\mu,\nu}\sigma c^{\dagger}_{i,\mu,\sigma}c_{i,\nu,\sigma}\nonumber\\           
     &+&\sum_{\mu,\nu}s^{+}_{\mu,\nu} c^{\dagger}_{i,\mu,\uparrow}c_{i,\nu,\downarrow}\nonumber\\           
     &+&\sum_{\mu,\nu}s^{-}_{\mu,\nu} c^{\dagger}_{i,\mu,\downarrow}c_{i,\nu,\uparrow}.
\end{eqnarray}     
Here $\mu,\nu=1,..,5$ is the index of the Fe $3d$ orbitals. We will use the following convention for the $3d$ orbitals, namely, $|1\rangle=|xz\rangle$, $|2\rangle=|yz\rangle$ and $|3\rangle=|xy\rangle$ for the three orbitals in the $t_{2g}$ subspace, $|4\rangle=|3z^{2}-r^{2}\rangle$ and $|5\rangle=|x^{2}-y^{2}\rangle$ for the two orbitals in the $e_{g}$ subspace. $\sigma=\uparrow,\downarrow$ is the spin of the valence electron. $c_{i,\mu,\sigma}$ is the annihilation operator for a spin-$\sigma$ electron on the $\mu$-th orbital of site $i$. The matrix elements in Eq.(5) are given by
\begin{eqnarray}
n_{\mu,\nu}&=&\frac{2}{3}\sum_{m}\langle \mu|r_{in}|m\rangle\langle m |r_{out}|\nu\rangle \nonumber\\
s^{z}_{\mu,\nu}&=&\frac{1}{3}\sum_{m} m\langle \mu|r_{in}|m\rangle\langle m |r_{out}|\nu\rangle \nonumber\\
s^{+}_{\mu,\nu}&=&\frac{1}{3}\sum_{m} A_{m}\langle \mu|r_{in}|m\rangle\langle m+1 |r_{out}|\nu\rangle \nonumber\\
s^{-}_{\mu,\nu}&=&\frac{1}{3}\sum_{m} A_{m}\langle \mu|r_{in}|m+1\rangle\langle m |r_{out}|\nu\rangle,
\end{eqnarray}     
in which $|m \rangle$ is the eigenstate of $l_{z}$ with eigenvalue $m$, $A_{m}=\sqrt{2-m(m+1)}$.

To simplify the analysis of the selection rule in the tetragonal environment, we expand the $2p$ core level $|m\rangle$ in the basis of real harmonics with the $p_{x}$, $p_{y}$ and $p_{z}$ character as
\begin{eqnarray}
|+1\rangle&=&-\frac{1}{\sqrt{2}}(|x\rangle+i|y\rangle)\nonumber\\
|0\rangle&=&|z\rangle\nonumber\\
|-1\rangle&=&\frac{1}{\sqrt{2}}(|x\rangle-i|y\rangle).
\end{eqnarray}     
One then find that the RIXS matrix elements become
\begin{eqnarray}
n_{\mu,\nu}&=&\frac{2}{3}\sum_{\alpha}\langle \mu|r_{in}|\alpha\rangle\langle \alpha |r_{out}|\nu\rangle \nonumber\\         
s^{\alpha}_{\mu,\nu}&=&\frac{i}{3}\sum_{\beta,\gamma}\epsilon_{\alpha\beta\gamma} \langle \mu|r_{in}|\beta\rangle\langle \gamma |r_{out}|\nu\rangle. 
\end{eqnarray}     
Here $\alpha=x,y,z$, $\epsilon_{\alpha\beta\gamma}$ is the anti-symmetric tensor.  

With these matrix elements in hand, we can calculate the RIXS cross section of Fe $3d$ excitation in both the density and the spin channel. In the paramagnetic phase, in which the spin is a conserved quantity, they are given by 
\begin{eqnarray}
I^{density}_{RIXS}(\mathrm{q},\omega)&\propto&-\mathrm{Im} [ \sum_{\mu\nu,\mu'\nu'}n^{*}_{\mu\nu}n_{\mu'\nu'}\  D_{\mu\nu,\mu'\nu'}(\mathrm{q},\omega) ]\nonumber\\
I^{spin}_{RIXS}(\mathrm{q},\omega)&\propto&-\mathrm{Im} [ \sum_{\mu\nu,\mu'\nu',\alpha\beta}(s^{\alpha}_{\mu\nu})^{*}s^{\beta}_{\mu'\nu'}\  \chi^{\alpha\beta}_{\mu\nu,\mu'\nu'}(\mathrm{q},\omega)].\nonumber\\
\end{eqnarray}
Here $D$ and $\chi$ are density and spin correlation function defined as follows
\begin{eqnarray}
D_{\mu\nu,\mu'\nu'}(\mathrm{q},\tau)&=& -\langle\ \mathrm{T}_{\tau} \mathrm{n}_{\mu\nu}(\mathrm{q},\tau) \ \mathrm{n}_{\mu'\nu'}(\mathrm{-q},0)\ \rangle\nonumber\\
\chi^{\alpha\beta}_{\mu\nu,\mu'\nu'}(\mathrm{q},\tau)&=& -\langle\ \mathrm{T}_{\tau} \mathrm{s}^{\alpha}_{\mu\nu}(\mathrm{q},\tau) \ \mathrm{s}^{\beta}_{\mu'\nu'}(\mathrm{-q},0)\ \rangle,\nonumber\\
\end{eqnarray}
in which $\mathrm{n}_{i,\mu\nu}=\sum_{\sigma}c^{\dagger}_{i,\mu,\sigma} c_{i,\nu,\sigma}$ is the density operator on site $i$, $\mathrm{s}^{\alpha}_{i,\mu\nu}=\frac{1}{2}\sum_{\gamma,\gamma'}c^{\dagger}_{i,\mu,\gamma}\sigma^{\alpha}_{\gamma,\gamma'} c_{i,\nu,\gamma'}$ is the spin density operator in the $\alpha$ direction on site $i$. In the magnetic ordered phase with a collinear AF order along the $z$ axis, the density fluctuation is entangled with the longitudinal spin fluctuation. One should then defined separately the density operator for the up spin and the down spin component, whose fluctuations are entangled\cite{Tohyama}.
 
To see more clearly the selection rule in the RIXS process, we split $s^{\alpha}_{\mu\nu}$ into its hermitian and anti-hermitian part, which correspond to the spin density and spin current excitation in the orbital space. In this study, we will focus on the spin density excitation, since the spin current excitation is in general a manifestation of the itinerant nature of the electron motion and only contribute a weak continuum in the spectrum. It is easy to show that the hermitian part of $s^{\alpha}_{\mu\nu}$ is asymmetric with respect to the exchange between $r_{in}$ and $r_{out}$ and thus proportional to $\epsilon_{in}\times\epsilon_{out}$. As a comparison, we note that the hermitian of $n_{\mu,\nu}$ is symmetric with respect to the exchange between $r_{in}$ and $r_{out}$.

Such a difference in the symmetry property of the spin and density excitation can be used to separate them in the RIXS measurement. To detect the spin excitation, one can employ the $\sigma\rightarrow \pi'$ scattering geometry(see Fig.2). Here $\sigma$ denotes that the polarization of the incident photon is perpendicular to the scattering plane, $\pi'$ denotes that the polarization of the scattered photon is within the scattering plane. Such a scattering geometry has already been used in a recent RIXS study on an electron-doped cuprate superconductor\cite{edop}. On the other hand, to detect the density difference between different Fe $3d$ orbitals, for example, in the electronic nematic phase of the IBSs, one can employ the  $\sigma\rightarrow \sigma'$ scattering geometry.

Before discussing the application of these results to the study of the IBSs, we note one limitation of the Fe $L_{3}$-edge RIXS. Since the largest momentum transfer at the Fe $L_{3}$-edge (with a photon energy of about $710 \ \mathrm{eV}$) is only about $\frac{\pi}{2a}$\cite{lattice}, the Fe $L_{3}$-edge RIXS can only cover one quarter of the first Brillouin zone in the IBSs. In particular, the most important momentum for the stripy magnetic correlation in the IBSs, namely, $\mathrm{Q}=(\pi,0)$ or $\mathrm{Q'}=(0,\pi)$, is out of the reach of the Fe $L_{3}$-edge RIXS. In the magnetic ordered phase, $\mathrm{Q}$ is folded back to the $\Gamma$ point so that the spin fluctuation around $\mathrm{Q}$ becomes accessible to the Fe $L_{3}$-edge RIXS measurement. At the same time, local moment fluctuation in the paramagnetic phase is expected to be robust even for momentum far away from the underlying magnetic ordering wave vector. The Fe $L_{3}$-edge RIXS can thus be used to study the orbital character of the spin fluctuation in both the magnetic ordered phase and the paramagnetic phase. 

In a previous study\cite{Tohyama}, the spin fluctuation in the both the magnetic ordered phase and the paramagnetic phase of the IBSs are computed with RPA for a five-band itinerant model. The result is then used to compute the RIXS cross section at the Fe $L_{3}$-edge. It is claimed that the long wave length spin wave excitation in the magnetic ordered phase has dominant $3d_{xy}$ character. On the other hand, it is found that no well defined dispersive mode can be resolved around the $\Gamma$ point in the paramagnetic phase from such a calculation. However, it is well known that the RPA scheme is unsuitable for the description of the local moment physics in the paramagnetic phase. In particular, it predicts that the spin fluctuation spectrum around the $\Gamma$ point in the paramagnetic phase should be essentially unchanged by the RPA correction and be composed of particle-hole continuum. This has been found to be incorrect by recent experiment on heavily doped IBSs, in which robust dispersive magnon mode has been observed around the $\Gamma$ point in the paramagnetic phase\cite{Zhou}. At the same time, the power of polarization resolution of the RIXS technique has not been fully explored in this early study, in which only the polarization of the incident photon is specified. This greatly limit the possibility of resolving the orbital character of the spin fluctuation in the IBSs with the RIXS technique.

Here we first consider the spin wave excitation of the magnetic ordered phase of the IBSs. As a symmetry restoration mode in the spin space, the long wave length spin wave excitation is expected to inherit the orbital character of the ordered moment. In a multi-orbital system, the magnetic order parameter is in general a matrix in the orbital space, with its matrix element defined by $M_{\mu,\nu}=\frac{1}{2}\sum_{\sigma}\sigma\langle c^{\dagger}_{i,\mu,\sigma}c_{i,\nu,\sigma}\rangle$. The orbital character of the ordered moment is encoded in the eigenvalues and eigenvectors of this matrix. In the IBSs, the lattice symmetry around the Fe ions is broken to $D_{2}$ in the magnetic ordered phase and the symmetry allowed magnetic order parameter can only take the form of
\begin{eqnarray}
\mathrm{\mathbf{M}}=\left(\begin{array}{cccccc} M_{1,1} & 0 & 0&|&0&0 \\0 & M_{2,2}&0&|&0&0\\ 0&0&M_{3,3}&|&0&0\\--&--&--&--&--&--\\ 0&0&0&|&M_{4,4}&M_{4,5}\\0&0&0&|&M_{4,5}&M_{5,5}\end{array}\right).
\end{eqnarray}     
Thus the ordered moment has either pure $t_{2g}$ or pure $e_{g}$ character. Previous studies indicate that the magnetic moment in the IBSs is mainly contributed by the $t_{2g}$ orbitals\cite{Dagotto,Si}. At the same time, since both the OSMT and the OSSF happen in the $t_{2g}$ subspace, it is also the most interesting place to study the orbital character of the spin fluctuation in the IBSs. We will thus focus on the spin fluctuation in the $t_{2g}$ subspace in the following discussion.

The selection rule for the Fe $L_{3}$-edge RIXS process becomes extremely simple in the $t_{2g}$ subspace. If we approximate the Wannier orbitals with the corresponding atomic orbitals\cite{table}, one find that the spin transition matrix element within the $t_{2g}$ subspace is given by
\begin{equation}
s^{\alpha}_{\mu,\nu}=\frac{c}{4}[(\mathrm{e}_{\mu}\cdot(\epsilon_{in}\times\epsilon_{out}))(\mathrm{e}_{\nu}\cdot\mathrm{e}_{\alpha})+\mu\leftrightarrow\nu](1-2\delta_{\mu\nu}).
\end{equation}
Here $c$ is a constant. $\mathrm{e}_{\mu}$ is a unit vector and is defined as $\mathrm{e}_{xz}=\mathrm{e}_{y}$, $\mathrm{e}_{yz}=\mathrm{e}_{x}$ and $\mathrm{e}_{xy}=\mathrm{e}_{z}$. $\epsilon_{in}$ and $\epsilon_{out}$ are the polarization vector of the incident and outgoing photon. For completeness' sake, the spin transition matrix element within the $e_{g}$ subspace is given by
\begin{equation}
s^{\alpha}_{\mu,\nu}=-\frac{c}{6}[\mathrm{e}_{\alpha}\cdot(\epsilon_{in}\times\epsilon_{out})](1-2\vec{\mathrm{n}}_{\alpha}\cdot\vec{\tau}_{\mu\nu}).
\end{equation}
Here $\vec{\tau}$ is the Pauli matrix in the $e_{g}$ subspace, $\vec{\mathrm{n}}_{\alpha}$ is a unit vector and is defined as $\vec{\mathrm{n}}_{\alpha}=\cos\phi_{\alpha}\mathrm{e}_{z}+\sin\phi_{\alpha}\mathrm{e}_{x}$, with $\phi_{z}=0$, $\phi_{x}=-\frac{2\pi}{3}$ and $\phi_{y}=\frac{2\pi}{3}$.

\begin{figure}
\includegraphics[width=9cm]{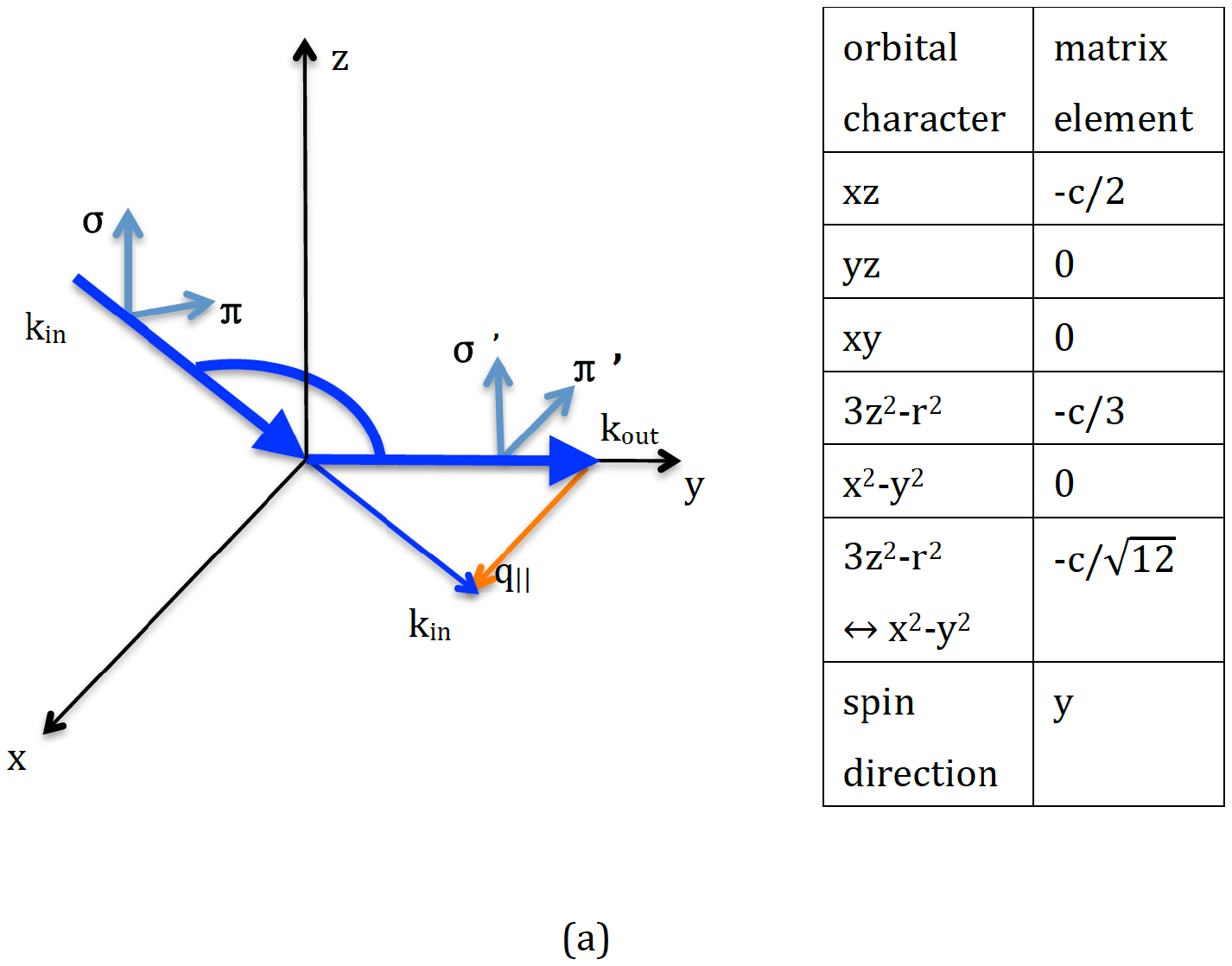}
\includegraphics[width=9cm]{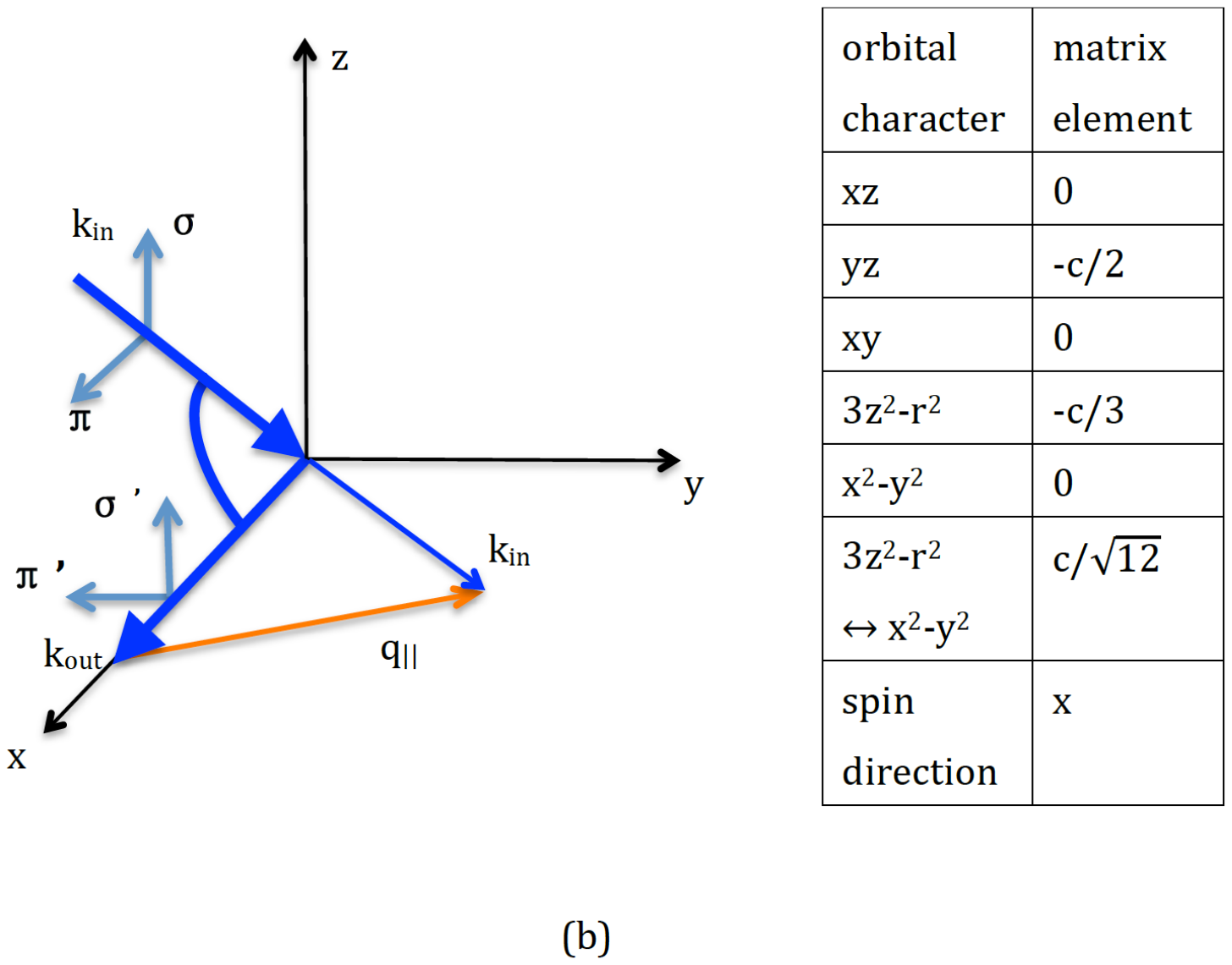}
\includegraphics[width=9cm]{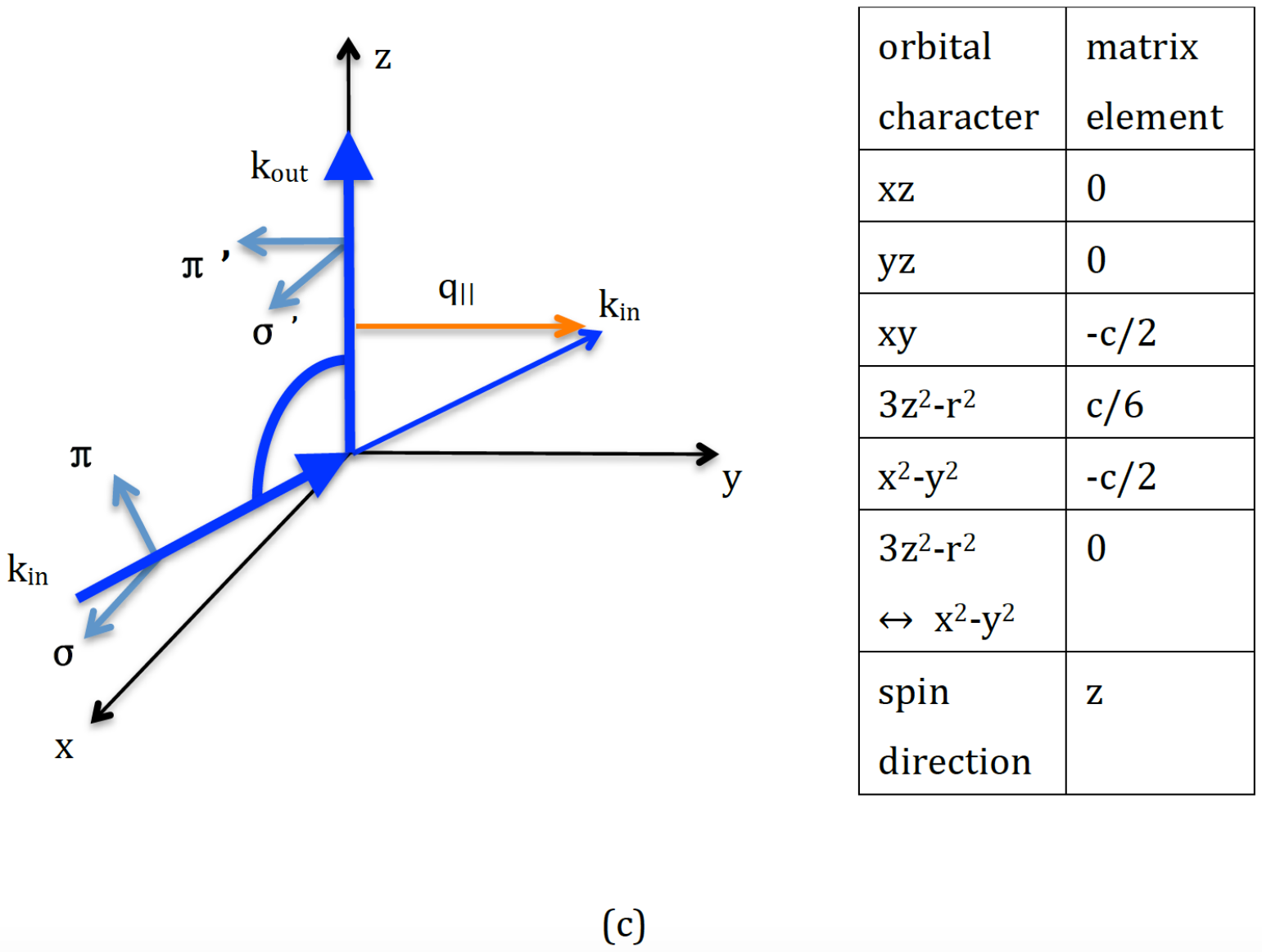}
\caption{\label{fig2}
(Color on-line) The RIXS scattering geometries proposed to detect the spin fluctuation on the (a)$3d_{xz}$, (b)$3d_{yz}$ and the (c)$3d_{xy}$ orbital and the corresponding RIXS matrix elements. Here we have adopted the $\sigma\rightarrow\pi'$ scattering geometry to maximize the matrix element of spin excitation. $\mathrm{k}_{out}$ is aligned with the $y$, $x$ and $z$-axis in (a), (b) and (c). The scattering plane has been set to be the $x-y$, $x-y$ and the $y-z$ plane in (a), (b) and (c) for clarity and can be rotated freely around $\mathrm{k}_{out}$. $\mathrm{q}_{||}$ is the momentum transfer in the $x-y$ plane and $c$ is a constant. The spin fluctuation excited in (a), (b) and (c) is along the $y$, $x$ and $z$ direction in the spin space. $3z^{2}-r^{2}\leftrightarrow x^{2}-y^{2}$ denotes the mixed $3z^{2}-r^{2}$ and $x^{2}-y^{2}$ orbital character.}
\end{figure}

We now discuss how to separate the spin fluctuation in the three Fe $t_{2g}$ orbitals in the RIXS measurement. First, since the RIXS matrix element for spin excitation is proportional to $\epsilon_{in}\times\epsilon_{out}$, the $\sigma\rightarrow\pi'$ scattering geometry is the most favorable for the detection of spin excitations. Second, according to Eq.(12), the orbital character of the spin fluctuation excited in the RIXS process is determined by the factor $\mathrm{e}_{\mu}\cdot(\epsilon_{in}\times\epsilon_{out})$. However, in the $\sigma\rightarrow\pi'$ scattering geometry, $\epsilon_{in}\times\epsilon_{out}$ is nothing but the unit vector in the direction of the outgoing photon. Thus, to measure the spin fluctuation in the $\mu$-th orbital, one should align the direction of the outgoing photon with $\mathrm{e}_{\mu}$. The excited spin fluctuation is also in the $\mathrm{e}_{\mu}$ direction in spin space. This is illustrated in Fig.2, in which we have listed the matrix elements for collective spin excitation in all major channels for the IBSs. Except for the uncertainty in the small spin fluctuation weight on the two $e_{g}$ orbitals, the orbital character of the spin fluctuation in the IBSs can thus be fully resolved by the proposed RIXS scattering geometries.

In the magnetic ordered phase of the IBSs, the ordered moment is found to lie in the FeAs plane and be perpendicular to the ordering wave vector\cite{order}. Assuming $\mathrm{Q}=(\pi,0)$, the ordered moment is then pointing in the $y$ direction. According to the selection rule listed above, one can either excite the transverse spin fluctuation in the $z$ direction in the $3d_{xy}$ orbital by aligning the direction of the outgoing photon with the $z$-axis, or excite the transverse spin fluctuation in the $x$ direction in the $3d_{yz}$ orbital by aligning the direction of the outgoing photon with the $x$-axis. The transverse spin fluctuation in the $3d_{xz}$ orbital is inaccessible with the Fe $L_{3}$-edge RIXS. One way to solve this problem is to apply a magnetic field in the $y$ direction so as to rotate the ordered moment to the $x$ direction or $z$ direction. In the paramagnetic phase, RIXS can be used to measure the spin fluctuation in all the three $t_{2g}$ orbitals with the same efficiency.

Such orbital-resolved information on the spin fluctuation can be used to address some important issues of the IBSs. The first issue is about the origin and nature of the OSMT in the IBSs. It has long been anticipated that electron on different Fe $3d$ orbitals may experience different strengths of electron correlation and that a two component picture with both itinerant electron and local moment is needed to understand the magnetism of the IBSs\cite{Zhang,Kou}. Among the five Fe $3d$ orbitals, it is generally believed that the $3d_{xy}$ orbital is the most strongly correlated\cite{Si,Shen}. Unlike the magnetism of the itinerant electrons, the local moment is expected to be robust against carrier doping and should survive even in the paramagnetic phase. Indeed, recent RIXS measurement does find evidence for the existence of dispersive magnon excitation in the heavily-doped IBSs\cite{Zhou}, whose dispersion is found to be very similar to the spin wave dispersion of the magnetic ordered phase of the parent compound. A natural question to ask is then if the observed local moment fluctuation in the heavily-doped IBSs has indeed a dominant $3d_{xy}$ orbital character. This problem can be answered by the RIXS measurement suggested above.

The second issue is about the origin and nature of the OSSF, or, the relation between orbital ordering and spin nematicity in the IBSs. Within the itinerant picture, it is straightforward to see that the electron on the $3d_{xz}$ and the $3d_{yz}$ orbital prefer different magnetic ordering patterns, since the electronic state around $(\pi,0)$ and $(0,\pi)$ have dominant $3d_{yz}$ and $3d_{xz}$ character. The orbital order that breaks the symmetry between the $3d_{xz}$ and $3d_{yz}$ orbital will thus be linearly coupled to the stripy magnetic order with wave vector $\mathrm{Q}$ or $\mathrm{Q'}$ and be generated spontaneously in the magnetic ordered phase\cite{Nevid,Su2,Fanfarillo1,Fanfarillo2,Fanfarillo3,Benfartto}. This scenario of OSSF can be verified by comparing the size of the ordered moment on the $3d_{xz}$ and the $3d_{yz}$ orbital in the magnetic ordered phase, which again can be inferred from the orbital character of the long wave length spin wave excitation. In principle, the orbital selectivity in the spin fluctuation pattern can also occur in the local moment scenario. To decide if this is the case, one can compare the spin fluctuation spectral weight in the $3d_{xz}$ and the $3d_{yz}$ orbital around the $\Gamma$ point in the paramagnetic phase, in which case the itinerant spin fluctuation becomes very weak.
 
Finally, we discuss the general structure of the spin fluctuation in the orbital space of the paramagnetic phase. For this purpose, we express the generalized spin susceptibility in terms of its spectral representation as follows
\begin{eqnarray}
\chi_{\mu\nu, \mu'\nu'}(\mathrm{q},i\omega_{n})=\int_{-\infty}^{\infty}\frac{d\omega'}{2\pi}\frac{R_{\mu\nu,\mu'\nu'}(\mathrm{q},\omega')}{i\omega_{n}-\omega'}.
\end{eqnarray} 
Here $R_{\mu\nu,\mu'\nu'}(\mathrm{q},\omega)$ denotes the matrix elements of the spectral density matrix and is given by
\begin{eqnarray}
R_{\mu\nu, \mu'\nu'}(\mathrm{q},\omega)&=& e^{\beta\Omega}\sum_{m,n}\langle n|\mathrm{s}_{\mu\nu}(\mathrm{q})|m\rangle\langle m|\mathrm{s}^{\dagger}_{\mu'\nu'}(\mathrm{-q})|n\rangle\nonumber\\
&\times& (e^{-\beta E_{n}}-e^{-\beta E_{m}})2\pi\delta(\omega+E_{n}-E_{m}),\nonumber\\
\end{eqnarray}
in which $E_{n}$ denotes the energy of the many-body eigenstate $|n\rangle$, $\Omega$ is the grand potential of the system. From this expression, it is straightforward to show that $R_{\mu\nu, \mu'\nu'}(\mathrm{q},\omega)$ is a positive definite hermitian matrix with respect to its index $\mu\nu$ and $\mu'\nu'$. The 25 eigenvalues $\rho^{(n)}$ and eigenvectors $V^{(n)}_{\mu,\nu}$ of $R$ can then be interpreted as the spectral weight and the orbital character of the spin fluctuation at momentum $\mathrm{q}$ and frequency $\omega$. In general, the eigenvectors $V^{(n)}_{\mu,\nu}(\mathrm{q},\omega)$ can have very complex momentum and frequency dependence and is not diagonal in the index $\mu$ and $\nu$. More specifically, it is composed of contributions from both the spin density and the spin current excitation in the orbital space. Such complications are manifestation of the itinerant nature of the spin fluctuation in the system, since the orbital character of local moment fluctuation is expected to be much more definite and much less sensitive to $\mathrm{q}$ and $\omega$. A measurement of the orbital character of the spin fluctuation in the paramagnetic phase can thus provide invaluable information on the electron correlation effect in the IBSs. We note that RIXS can be used to detect both the spin density and the spin current fluctuation in the orbital space.

In summary, we propose that the Fe $L_{3}$-edge RIXS can be used to detect the orbital character of spin fluctuations in the IBSs. In particular, we find that the spin fluctuation on the three $t_{2g}$ orbitals, namely, the $3d_{xz}$, $3d_{yz}$ and the $3d_{xy}$ orbital, can be selectively probed in the $\sigma\rightarrow\pi'$ scattering geometry by aligning the direction of the outgoing photon with the $y$, $x$ and $z$ axis. We show that such orbital-resolved information on the spin fluctuation can be very useful in the study of the OSMT and the OSSF physics in the IBSs and deepen our understanding on the origin of the electronic nematicity and pairing mechanism in the IBSs. The proposed technique can also be used in the study of other transition metal oxides in which the $t_{2g}$ orbital is dominating the low energy physics.

We acknowledge the support from the grant NSFC 11674391, the Research Funds of Renmin University of China and the grant National Basic research project 2016YFA0300504.


\begin{thebibliography}{99}
\bibitem{Zhang}G.M. Zhang, Y.H. Su, Z.Y. Weng, D.H. Lee, and T. Xiang, Euro. Phys. Lett. \textbf{86 }37006 (2009).
\bibitem{Kou}S.P. Kou, T. Li, and Z.Y. Weng, Euro. Phys. Lett. \textbf{88} 17010 (2009).
\bibitem{Shen}M. Yi, Z.K. Liu, Y. Zhang, R. Yu, J.X. Zhu, J. Lee, R. Moore, F. Schmitt, W. Li, S. Riggs, J.H. Chu, B. Lv, J. Hu, M. Hashimoto, S.K. Mo, Z. Hussain, Z.Q. Mao, C.W. Chu, I. Fisher, Q.M. Si and Z.X. Shen and D.H. Lu, Nat. Comm. \textbf{6}, 7777 (2015).
\bibitem{Si}R. Yu, Q.M. Si, Phys. Rev. Lett. \textbf{110}, 146402 (2013).
\bibitem{Capone}L. Medici, G. Giovannetti and M. Capone, Phys. Rev. Lett. \textbf{112}, 177001 (2014).
\bibitem{Lee}C.C. Lee, W.G. Yin and W. Ku, Phys. Rev. Lett. \textbf{103}, 267001 (2009).
\bibitem{Dagotto}M. Daghofer, Q.L. Luo, R. Yu, D. X. Yao, A. Moreo and E. Dagotto, Phys. Rev. B \textbf{81}, 180514 (2010).
\bibitem{Nevid}A. H. Nevidomskyy, arXiv.org:1104.1747 (2011).
\bibitem{Su1}Y.H. Su, H.J. Liao and T. Li, J. Phys.: C \textbf{27} 105702(2015).
\bibitem{Su2}Y.H. Su, C. Zhang and T. Li, Phys. Lett. A \textbf{380} 2008(2016).
\bibitem{Fanfarillo1}L. Fanfarillo, A. Cortijo, and B. Valenzuela, Phys. Rev.B, \textbf{91},214515(2015).
\bibitem{Fanfarillo2}L. Fanfarillo, L. Benfatto, and B. Valenzuela, Phys. Rev.B, \textbf{97},121109(R)(2018).
\bibitem{Fanfarillo3}L. Fanfarillo, J. Mansart, P. Toulemonde, H. Cercellier, P. Le F\`evre, F. Bertran, B. Valenzuela, L. Benfatto, and V. Brouet, Phys. Rev.B, \textbf{94},155138(2016).
\bibitem{Benfartto}L. Benfartto, B. Valenzuela and L. Fanfarillo, arXiv:1804.05800.
\bibitem{Beak}S.H. Baek, D.V. Efremov, J.M. Ok, J.S. Kim, J. van den Brink, B. B\"uchner, Nat. Mater., \textbf{14}, 210 (2015).
\bibitem{Luo}H.Q. Luo, M. Wang, C.L. Zhang, X.Y. Lu, L.P. Regnault, R. Zhang, S.L. Li, J.P. Hu, and P.C. Dai, Phys. Rev. Lett. \textbf{111}, 107006 (2013).
\bibitem{Kuroki}K. Kuroki, S. Onari, R. Arita, H. Usui, Y. Tanaka, H. Kontani, and H. Aoki, Phys. Rev. Lett.\textbf{101}, 087004 (2008).
\bibitem{LeeDH}Y. Ran, F. Wang, H. Zhai, A. Vishwanath and D.H. Lee, Phys. Rev. B \textbf{79}, 014505 (2009).
\bibitem{Kotliar}Z.P. Yin, K. Haule and G. Kotliar, Nat. Phys. \textbf{10}, 845(2014).
\bibitem{Kreisel}A. Kreisel, B. M. Andersen, P. O. Sprau, A. Kostin, J. C. S\'eamus Davis, and P. J. Hirschfeld, Phys. Rev.B \textbf{95}, 174504(2017).
\bibitem{Sprau}P. O. Sprau, A. Kostin, A. Kreisel, A. E. B\"ohmer, V. Taufour, P. C. Canfield, S. Mukherjee, P. J. Hirschfeld, B. M. Andersen, J. C. S\'eamus Davis, Science \textbf{357},75 (2017).

\bibitem{RIXS0}L.J.P. Ament, G. Ghiringhelli, M. M. Sala, L. Braicovich and J. van den Brink, Phys. Rev. Lett. \textbf{103}, 117003 (2009).
\bibitem{RIXS}L.J.P. Ament, M. van Veenendaal, T.P. Devereaux, J.P. Hill, and J. van den Brink, Rev. Mod. Phys. \textbf{83}, 705 (2011).
\bibitem{Zhou}K.J. Zhou, Y.B. Huang, C. Monney, X. Dai, V. N. Strocov, N.L. Wang, Z.G. Chen, C. Zhang, P. Dai, L. Patthey, J. van den Brink, H. Ding, and T. Schmitt, Nat Comm \textbf{4}, 1470 (2013).
\bibitem{Tohyama}E. Kaneshita, K. Tsutsui and T. Tohyama, Phys. Rev. B \textbf{84}, 020511(\textrm{R}) (2011).

\bibitem{LuoJ}J. Luo, G. T. Trammell, and J. P. Hannon,  Phys. Rev. Lett. \textbf{71}, 287(1993).
\bibitem{Veenendaal}M. van Veenendaal,  Phys. Rev. Lett. \textbf{96}, 117404(2006).
\bibitem{indirect}For indirect RIXS processes, the coupling between the valence electron and the core hole can have important consequences. 

\bibitem{edop}E.H. da Silva Neto, M. Minola, B. Yu, W. Tabis, M. Bluschke, D. Unruh, H. Suzuki, Y. Li, G. Yu, D. Betto, K. Kummer, F. Yakhou, N.B. Brookes, M. LeTacon, M. Greven, B. Keimer, and A. Damascelli, arXiv:1804.09185.

\bibitem{lattice}$a\approx4\ \AA$ is the lattice constant of the IBSs.

\bibitem{order}C. de la Cruz, Q. Huang, J. W. Lynn, J. Li, W. Ratcliff II, J. L. Zarestky, H. A. Mook, G. F. Chen, J. L. Luo, N. L. Wang and P.C. Dai, Nature \textbf{453}, 899 (2008).

\bibitem{table}The RIXS selection rule for spin excitation in the $t_{2g}$ subspace discussed in the main text is essentially unchanged even if we do not approximate the Wannier orbitals with the corresponding atomic orbitals. In fact, using the fact that the three $t_{2g}$ orbitals and the three $2p$ orbitals each form a distinct one dimensional representation of the $D_{2}$ group, it is easy to obtain the following selection rules.
\begin{tabular}{|c|c|c|c|c|}
\hline
orbital & $r_{in}$ & $r_{out}$ & spin direction &  transition probability \\
\hline
xz & x(z) & z(x)  & y & $|\langle xz|x|z\rangle|^{2}$ \\
\hline
yz & y(z) & z(y)  & x & $|\langle yz|y|z\rangle|^{2}$ \\
\hline
xy & x(y) & y(x)  & z & $|\langle xy|x|y\rangle|^{2}$ \\
\hline
\end{tabular}


\end{thebibliography}
\end{document}